# Load Estimation for Electric Power Distribution Networks

***Chiebuka Eyisi, Saeed Lotffifard***
Department of Electrical Engineering and Computer Science
University of Central Florida
Orlando, U.S.A

**Abstract:** *Distribution Load Estimation (DLE) is a key function of Distribution Management System (DMS). In this paper a novel method for presenting historical load data in the form of Representative Load Curves (RLC) is presented. Adaptive Neuro-Fuzzy Inference Systems (ANFIS) is used in this regard to estimate the RLC. Accurate RLCs provide better pseudo-measurements for real-time load estimation in distribution networks. The performance of the proposed method is demonstrated on an 11kV radial distribution network with the aid of the MATLAB software. The mean absolute percent error (MAPE) criterion is used to quantify the accuracy of the estimated RLC.*

## Introduction

In the electricity industry, the major determinant in electricity supply strategy is the quantity of demand. Direct measurements and historical measured data are two reliable sources of data commonly utilized in several distribution load estimation (DLE) techniques. Distribution substations, transformers, and some important loads are usually the only sources of direct measurements. With the aid of historical data, one can generate more accurate pseudo-measurements in the form of estimate representative load curves (RLC) to help augment unavailable measurements in a given DLE process. An RLC is a curve representing load's aggregate behavior over a particular period or specified time span. They are usually built in the set of similar load curves.

For efficient operation and control of power distribution networks, updated nodal load information is needed to represent customers accurately. How electric energy is used at various times and the aggregated customers' share of the utility's total load is of major emphasis to Distribution System Operators (DSO). Load estimates are needed in advanced functions of Distribution Automation (DA). Historical data is usually mined for the abstraction of load information from billing cycle customer kWh consumption and major distribution substations and equipment installations and used to augment unavailable measurements. Estimate RLCs are obtained from this data as presented in this paper.

## Related Work

An overview of DLE is presented in [1]. In [2], [3], load estimates were attained by scaling measurements according to transformer's peak load analysis. In [4], actual load curves were used to scale measurements to obtain load estimates. In [5], [6], the idea using distribution state estimation (DSE) techniques for DLE is proposed. In [7], a two-step procedure combining load allocation with DSE techniques is proposed. A statistical load modeling method to express the variation of active power demand in radial networks is proposed in [8]. In [9], fuzzy regression is utilized in radial networks to express the correlation between substation peak active loads and supplied customer active loads. In [10], expert knowledge and operator experience are utilized, where fuzzy variables were used to represent linguistic descriptions for the size of loads. In [11], a Weighted Least Absolute Value (WLAV) approach is proposed to reduce the effect of gross errors in measurements. In [12], neural and fuzzy set techniques are used to obtain load curves for customers' classes based on their monthly energy consumption and a large set load curves' data obtained from measurements. A zonal load estimation that divides radial distribution networks into several zones is proposed in [1], [13], [14]. In [15], a Weighted Least Squares (WLS) method is proposed for DLE.

A Case-Based-Reasoning (CBR) method is presented for distribution network nodal load estimation in [16]. A DLE method using clustering techniques is presented in [17] in which k-means clustering is used as part of the estimation process to obtain coarse estimates and these estimates are further refined to obtain the estimated load. In [18] Fuzzy Inference Systems (FIS) are utilized to estimate substation load by aggregating individual FIS of Takagi-Sugeno type. Pseudo-measurements play a significant role in the accuracy of the load estimates. Some choices include; average daily customer demand and/or classified typical load curves. RLCs are customer/node dependent and are obtained from historical data.

Highly accurate Representative Load Curves can be used to describe electricity demand. In [19], Multiple Discriminant Analysis (MDA) is used to cluster daily load curves into a set of RLCs. It's important that RLCs capture the dynamics of demand variations. An order-specific clustering algorithm for the determination of representative load curves is proposed in [20]. In [21], RLCs were built on the basis of their electricity behavior utilizing Fuzzy K-Means (FKM). In [22], FKM was also applied to determine typical load profiles of consumers. An algorithm utilizing FKM and hierarchical clustering for determining RLCs is also presented in [23]. In this paper, a method utilizing ANFIS is proposed to estimate RLCs from historical measured data.

## Anfis

ANFIS integrates the best features of Fuzzy Systems (FS) and Artificial Neural Networks (ANN) and involves utilizing linguistic information from the FS as well as the learning capability of an ANN for automatic fuzzy IF-THEN rule generation and parameter optimization [24]. ANFIS comes into play when one can't ascertain what membership functions look like simply from looking at data. With ANFIS, one can construct a FIS whose membership function parameters are tailored using a *hybrid algorithm* (back-propagation and least-squares), on a well-represented input/output dataset in order to account for variations in the data values. The ANFIS architecture consisting of two fuzzy IF-THEN rules based on Takagi and Sugeno's type is as follows [25].

Rule 1: if $x$ is $A_1$ and $y$ is $B_1$, then $f_1 = p_1x + q_1y + r_1$.
Rule 2: if $x$ is $A_2$ and $y$ is $B_2$, then $f_2 = p_2x + q_2y + r_2$.

Each layer contains nodes (neurons) described by the same function. In summary, when the values of the premise parameters are learned, the overall output can be expressed as a linear combination of consequent parameters [24].

## Methodology

A *Time Series Estimation* was attempted heuristically on the input/target dataset to estimate RLCs. Both sets of data must be within the load profile. Various adaptations for the input dataset could include: present load ($t$), previous 24 hour load ($t$ – 24), previous week same hour load ($t$ – 168), weekly model factor (WMF), daily model factor (DMF). The model factors are computed using historical substation or nodal data by dividing the average load in its specific week or day by the maximum load in that same select week or day. These adaptations are chosen heuristically for different cases with the sole objective of seeking a model that generates the least Minimum Absolute Percent Error (MAPE). 4 weeks in each month are heuristically selected to be used to determine some sort of Weekly and Daily RLCs for each month with regards to the overall system load (RTD) or nodal demand. This was attained with the aid of the MATLAB software [26] as somewhat of a data-fitting process. A daily and/or weekly RLC was chosen on one basic criterion:

*The Mean Absolute Percent Error (MAPE) between the corresponding sections/windows (between the selected estimated RLC and the target vector) must be the least.*

For instance, the utilized historical load data in this paper is for a 2-year period. In March, only 8 weeks are to be used in this analysis (first 4 weeks in 2009 & 2010). In selecting the weekly RLC, from the estimated RLC (fitted ANFIS output), the estimated RLC and the corresponding target vector are divided into eight sections/windows to represent the selected eight weeks (1344 hours' samples of data). Hence there are two sets of data broken into 168 samples. That's eight sections/parts with 168 hours' samples each.

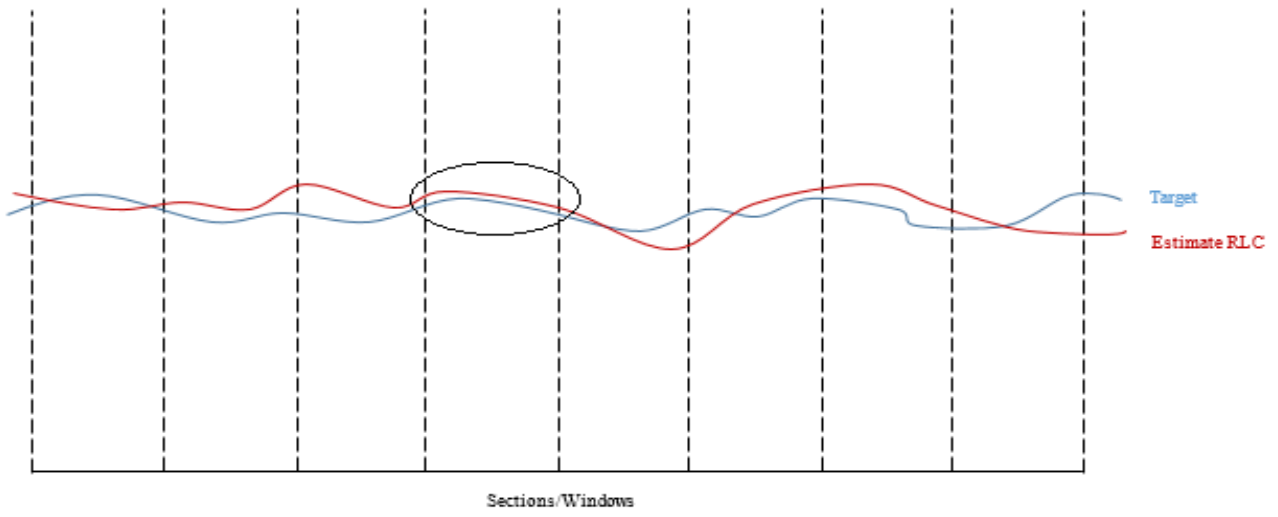

**Figure 1.** Selected Estimate RLC Description

Comparing each section, the method selects the section of the estimated RLC that has the least MAPE when compared with the corresponding target data. This selection of the fitted ANFIS output then represents the weekly RLC. A pictorial description is shown in fig. 1. In this case the encircled fourth window is therefore selected. Same analogy can be applied in selecting the daily RLC using the same 8 weeks in this case. Corresponding days (Mondays, Tuesdays, etc) of the target vector are grouped together and the daily RLC is selected again upon the basic MAPE criterion. Heuristically, after trying different adaptations of the respective target output, the models summarized below achieved the least overall MAPE when estimating weekly/daily RLCs to represent monthly historical data for the overall system load and/or select node; whatever the case may be. For instance, Y($t$) would represent the present load in the load profile used in our case study and is one of the inputs to the ANFIS as described below. This input/target pair must be within the specified load profile. The input/target pair to the ANFIS are as follows:

*Weekly/Daily RLC for Overall System Load:*
Inputs (X): WSMF/DSMF, Y($t$ – 168), (Y($t$ – 24)), Y($t$). Target (T): Y($t$ + 24). WSMF/DSMF represent weekly system model factor and daily system model factor respectively.

*Weekly/Daily RLC for Select Node:*
Inputs (X): WNMF/DNMF, Y($t$ – 168), Y($t$ – 24), Y($t$). Target (T): Y($t$ + 24). WNMF/DNMF represent weekly nodal model factor and daily nodal model factor respectively.

The summarized modeling criterion used for both cases in the MATLAB software is shown in table I. In both cases, the input/target dataset is split into two (training and checking). Then an initial FIS is generated with the training dataset and validated with the checking dataset. Also, because the previous week's data is needed as one of our selected inputs, the estimate RLCs obtained from the ANFIS fit excluded the 1st week's data (168 samples) in the estimation/data-fitting process. This entry is replaced with NaN.

**TABLE 1**
MODELING CRITERION FOR ESTIMATING RLCS

| *S/N* | *Custom ANFIS* | *Variables* |
|---|---|---|
| 1 | Number of Inputs | Four |
| 2 | Membership Function Type | Generalized Bell |
| 3 | Number of Membership Functions | Varied from 2 to 3 |
| 4 | Learning Algorithm | Hybrid Learning Algorithm |
| 5 | Epoch Size | Varied from 50 to 100 |
| 6 | Data Size | Data per Hour (17520 Samples) |
| 7 | Sugeno-Type System | First Order |
| 8 | Output Type | Linear |
| 9 | Initial Step Increase/Decrease Size | 1.1/0.9 |
| 10 | Number of Linear Parameters | 80 |
| 11 | Number of Nonlinear Parameters | 24 |
| 12 | Number of Nodes | 55 |
| 13 | Number of Fuzzy Rules | 16 |
| 14 | Data/Parameter Ratio | ≈ 84 |

The discussed method is summarized using the following steps:

1) Heuristically organize input/output dataset.
2) Initialize the fuzzy system using pre-defined MATLAB functions.
3) Give other parameters for ANFIS learning: Error (Tolerance), Number of Epochs.
4) Begin ANFIS learning using MATLAB *anfis* command. Stop when the number of epochs is reached or tolerance is obtained.
5) Validate with independent data adressing any over-fitting concerns.
6) Use defined MAPE criterion to obtain estimate RLC if step 5 above is confirmed. Else go to step 1 and repeat process.

## Case Study and Results

The proposed method for estimating RLCs is implemented on a sample 11kV radial distribution feeder employed from [28]. System loads are considered as constant (PQ) power or spot loads. The load data used was mined and extracted data from [29] and was adapted to suit the test network. This dataset comprised two years of historical hourly measured data (Jan 1st 2009 – Dec 31st 2010).

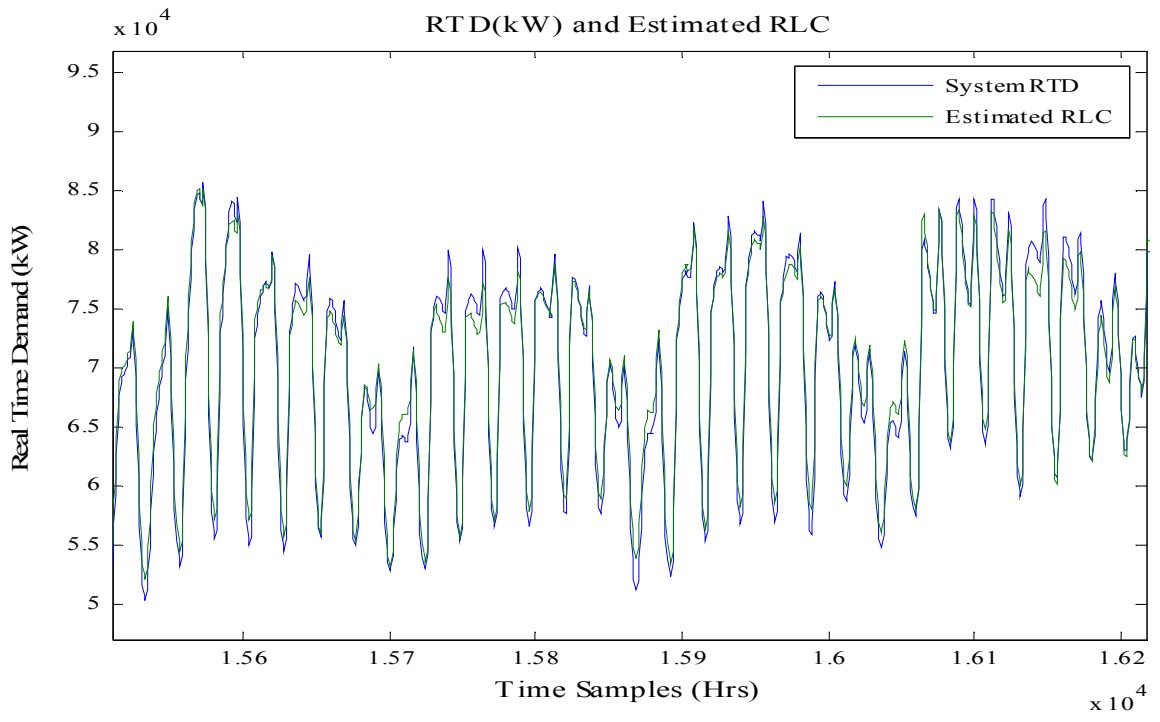

**Figure 2.** Magnified Estimate RLC & System RTD (for Weekly RLC)

The ANFIS structure follows the modeling criterion in table 1 for both cases explained in the methodology. The results upon training/testing the ANFIS of the estimated RLC and system RTD, intended for weekly estimate RLC are further shown in fig. 2 with overall MAPE of 2.5071% and root mean square error, RMSE of $6.2790 \times 10^3$ kW. Similar results can be obtained for a selected node with similar adaptations. The initial and adjusted input membership functions for this estimation are shown in the figs. 3 and 4 respectively.

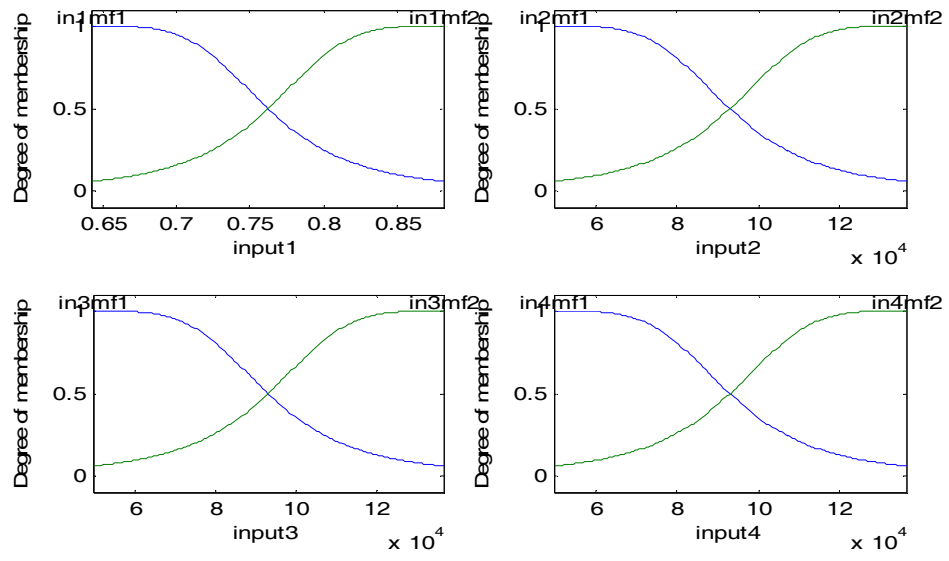

**Figure 3.** Initial Input Memebership Functions

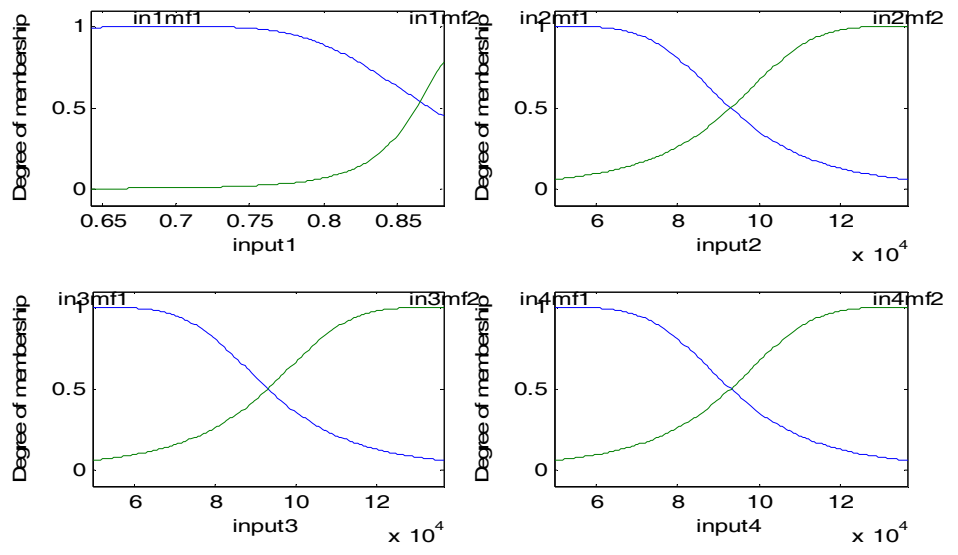

**Figure 4.** Adjusted Input Membership Functions

The error between system load and estimate RLC (intended for weekly RLC) in kW is shown in fig. 5.

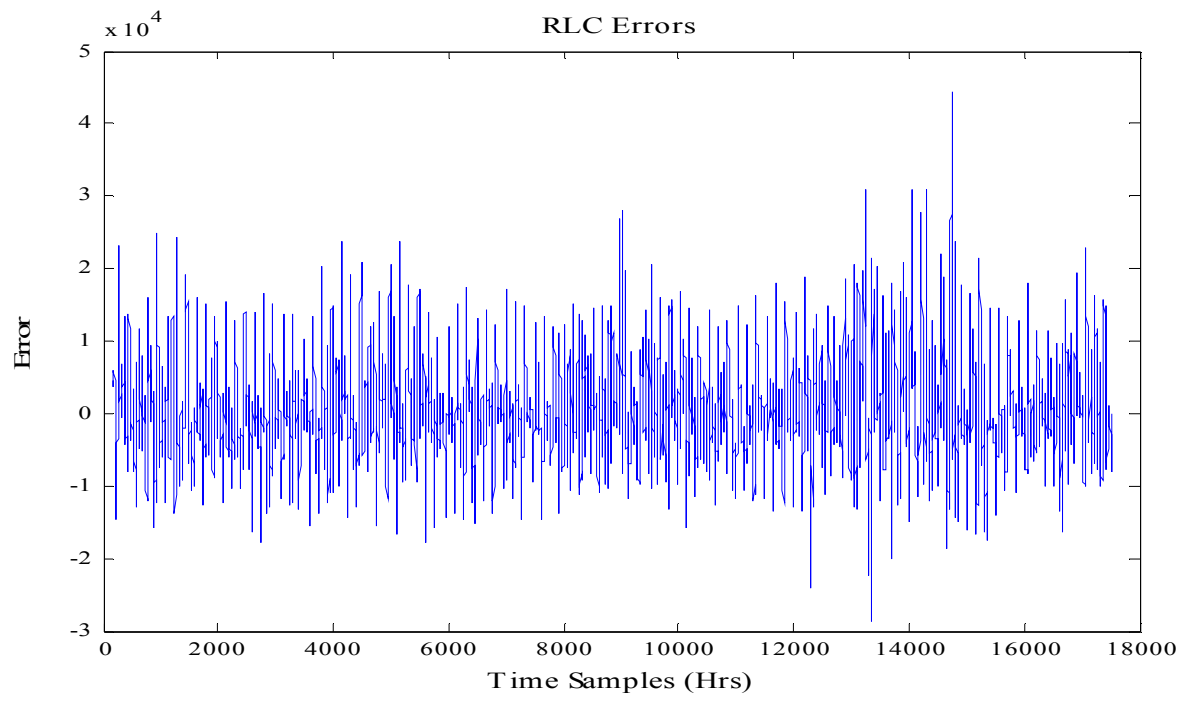

**Figure 5.** Error between System Load & Estimate RLC (for Weekly RLC)

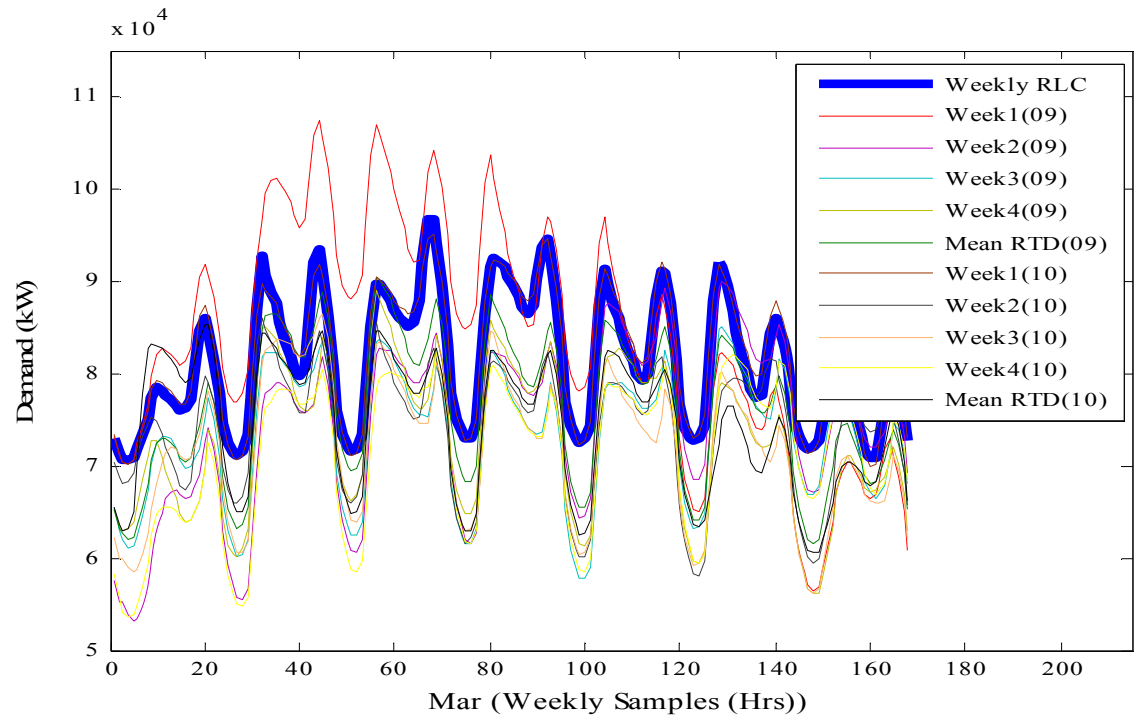

**Figure 6.** System, March Estimate Weekly RLC

The fig. 6 and table 2 below illustrates how the estimate weekly RLC for the month of March was selected using the MAPE criterion. Similar results can be obtained when estimating weekly/daily RLCs for a select node. Results of estimated daily nodal RLC for a select node 5 [28] for example were obtained with overall MAPE of 3.3523% and RMSE of 363.024kW.

**TABLE 2**
TABULAR MAPE SHOWING WEEKLY COMPARISONS

| | ***Year 2009*** | | | | ***Year 2010*** | | | |
|---|---|---|---|---|---|---|---|---|
| | ***January*** | | | | | | | |
| Window/ Week | 1 | 2 | 3 | 4 | 1 | 2 | 3 | 4 |
| Estimate RLC MAPE | NaN | 3.04 39 | 3.65 80 | 2.06 72 | 3.52 47 | 4.26 99 | **1.98 37** | 3.45 45 |

| | *Year 2009* | | | | *Year 2010* | | | |
|---|---|---|---|---|---|---|---|---|
| | ***February*** | | | | | | | |
| Window/ Week | 1 | 2 | 3 | 4 | 1 | 2 | 3 | 4 |
| Estimate RLC MAPE | 2.9859 | 2.1857 | **1.2555** | 1.8272 | 3.5504 | 2.4485 | 3.5648 | 1.9084 |
| | ***March*** | | | | | | | |
| Window/ Week | 1 | 2 | 3 | 4 | 1 | 2 | 3 | 4 |
| Estimate RLC MAPE | 2.7775 | 2.9444 | 1.4536 | 1.9165 | **1.2002** | 2.7785 | 1.9662 | 2.5529 |

## Conclusions

A systematic approach using ANFIS is developed to construct a Representative Load Curve from historical data. The present load, previous 24 hour load, previous week same hour load and weekly system model factors or daily system model factors are used as inputs to obtain the best representative curves for a select week or day as the case may be. Least Minimum Absolute Percent Error (MAPE) is defined as a criteria to quantify the accuracy of the estimates. The proposed algorithm was implemented on an 11kV network and it is demonstrated that the accuracy of the estimates were within 2-4%. Although the proposed algorithm was demonstrated for substation data, it can be applied to nodal load data as well. The constructed RLC can be utilized for real-time distribution load estimation.

## References

[1] Jie Wan, "Nodal load estimation for electric power distribution systems", *Ph.D. Dissertation*, Drexel University, 2003.

[2] V. Borozan and N. Rajakovic, "Load estimation for distribution systems with minimum information," *Euro. Trans. On Electr. Power*, iss. 14, pp. 331-345, Aug. 2004.

[3] V. Borozan and N. Rajakovic, "Minimum loss distribution network configuration: analyses and management," Electricity Distribution. Part 1: Contributions. CIRED. *14th International Conference and Exhibition on IEE Conference*, no. 438, vol. 6, 1997, pp. 18/1 - 18/5.

[4] R. P. Broadwater, A.H. Khan, H. E. Shaalan and R. E. Lee, "Time varying load analysis to reduce distribution losses through reconfiguration," *IEEE Trans. Power Delivery*, vol. 8, no. 1, pp. 294-300, Jan. 1993.

[5] M. E. Baran and A. W. Kelley, "State estimation for real-time monitoring of distribution systems," *IEEE Trans. Power Syst.*, vol. 9, no. 3, pp. 1601-1609, Aug. 1994.

[6] M. E. Baran, "Load estimation for distribution system analysis," Panel Summary, *Proceedings of the 2001 IEEE Power Engineering Society Summer Meeting*, Vancouver, Canada.

[7] T. Wang and M. Fan, "A novel load estimation method in distribution networks", in *Proc. Power Systems Technology*, vol. 1.1, 1998, pp. 567-571.

[8] A. K. Ghosh, S. L. Lubkeman and R. H. Jones, "Load modeling for distribution circuit state estimation," *IEEE Trans. Power Delivery*, vol. 12, no. 2, pp. 999-1005, Apr. 1996.

[9] J. Nazarko and W. Zalewski, "The fuzzy regression approach to peak load estimation in power distribution systems", *IEEE Trans. on Power Systems*, vol. 14, no.3, pp. 809-814, Aug. 1999.

[10] H. Kuo and Y. Hsu, "Distribution system load estimation and service restoration using a fuzzy approach", *IEEE Trans. On Power Delivery*. vol. 8, no. 4, Oct. 1993.

[11] M.R. Irving and C.N. Macqueen, "Robust algorithm for load estimation in distribution networks", IEE Proc.-Gener. Transm. Distri., vol. 145, no. 5, pp. 499-504, Sept. 1998.

[12] D. M. Falcao and H. O. Henriques, "Load estimation in radial distribution systems using neural networks and fuzzy set techniques," Panel Summary, *Proceedings of the 2001 IEEE Power Engineering Society Summer Meeting*, Vancouver, Canada.

[13] J. Wan, K.N. Miu, "A zonal-load estimation method for unbalanced radial distribution networks", *IEEE Trans. on Power Delivery*, vol. 17, no. 4, pp. 1106-1112, Oct. 2002.

[14] J. Wan, K.N. Miu, "Load estimation in radial electric power distribution networks using limited measurements", in *Proc. IEEE International Symposium on Circuits and Systems*, vol.2, Geneva, Switzerland, 2000, pp. 517-520.

[15] J. Wan, K.N. Miu, "A WLS Method for load estimation in unbalanced distribution networks", *IEEE Power Engineering Society Winter Meeting*, vol. 2, pp. 969-974, Jan. 2002.

[16] J. Wu, Y. Yu, "CBR-based load Estimation for distribution networks", in *Proc. IEEE Electrotechnical Conference*, MELECON, Malaga, Spain, May 2006, pp. 952-955.

[17] G. Grigoraş, F. Scarlatache and G. Cârţină, "Load estimation for distribution systems using clustering techniques", *Optimization of Electrical and Electronic Equipment (OPTIM), 2012 13th International Conference*, pp. 301 – 306, May 2012.

[18] T. Konjic, V. Miranda and I. Kapetanovic, "Fuzzy inference systems applied to LV substation load estimation", *IEEE. Trans. On Power Sys.*, vol. 20, no. 2, pp. 742-749, May 2005.

[19] P. Balachandra and V. Chandru, "Modelling electricity demand with representative load curves", Elsevier Energy, vol. 24, iss. 3, pp. 219-230, Mar. 1999.

[20] C.H. Marton, A. Elkanel, T.A. Duever, "An order-specific clustering algorithm for the determination of representative demand curves", Elsevier Computer and Chemical Engineering, vol. 32, pp. 1365-1372, Jun. 2007.

[21] P.T.T. Binh, N.H. Ha, T.C. Tuan and L.D.Khoa, "Determination of representative load curve based on Fuzzy K-Means" in *Proc. 4th Intl. Power Engineering and Optimization Conf (PEOCO2010)*, Malaysia, pp. 281-286, Jun. 2010.

[22] J. Hossain, A.N.M. Enamul Kabir, M. Rahman, B. Kabir and R. Islam, "Determination of typical load profile of consumers using Fuzzy C-Means clustering algorithm", *International Journal of Soft Computing and Engineering (IJSCE)*, vol. 1, iss. 5, pp. 169-173, Nov. 2011.

[23] D. Gerbec, S. Gasperic, I. Simon and F. Gubina, "Consumer's load profile determination based on different classification methods", in *Proc. Power Engineering Society General Meeting*, vol. 2, pp. 900-905, Jul. 2003.

[24] J.S. Jang, "ANFIS: adaptive-network-based fuzzy inference systems", *IEEE Trans. Sys. Man. Cybern.*, vol. 23, Iss. 3, pp. 665-685, May/Jun 1993.

[25] T. Takagi and M. Sugeno, "Derivation of fuzzy control rules from human operator's control actions", in *Proc. IFAC Symp Fuzzy Inform, Knowledge Representation and Decision Analysis*, pp. 55-60.

[26] Fuzzy Logic Toolbox™ User's Guide. www.mathworks.com Available [Online].

[27] J.-S. R. Jang, "Input selection for ANFIS learning", *in Proc. 5th IEEE Intl. Conf. on Fuzzy Systems*, vol. 2, pp.1493-1499, Sep. 1996.

[28] D. Das, D.P. Kothari and A. Kalam, "Simple and efficient method for load flow solution of radial distribution networks", Elsevier Electric Power and Energy Systems, vol. 17, no. 5, pp. 335-346, 1995.

[29] Historical Metered Load. www.pjm.com/markets-and-operations/ops-analysis/historical-load-data.aspx Available [Online].

[30] W.H. Kersting, *Distribution System Modeling and Analysis*, CRC Press Inc, New Mexico, 2002.

[31] J.M. Zurada, Introduction to Artificial Neural Systems, West Publishing Company, Minnesota, 1992

[32] A. Kandel, *Fuzzy Expert Systems*, CRC Press Inc, Boca Raton, Florida, 1992.